\documentclass[letterpaper, 10 pt, conference]{ieeeconf}
\IEEEoverridecommandlockouts  
\let\labelindent\relax
\usepackage{enumitem}

\usepackage{mathtools}
\usepackage{url}
\usepackage{hyperref}
\usepackage{graphics} 
\usepackage{epsfig}
\usepackage{cite}
\usepackage{times} 
\usepackage{amsmath,amssymb,amsfonts}
\usepackage{algorithm}
\usepackage{algpseudocode}
\usepackage{graphicx}
\usepackage{textcomp}
\usepackage{varioref}
\usepackage{import}
\usepackage{framed,xcolor}
\usepackage{color}
\usepackage{comment}
\usepackage{epstopdf}   
\usepackage{psfrag}
\usepackage{breqn}
\usepackage{float}
\usepackage{booktabs}
\usepackage{soul}
\usepackage{amsbsy}
\usepackage[bottom]{footmisc}
\usepackage{lineno}
\usepackage{cleveref}
\usepackage{microtype}
\usepackage{comment}
\usepackage{arydshln}
\usepackage{mathtools}
\usepackage{optidef}
\usepackage{graphicx}

\usepackage{tikz}
\usepackage{standalone} 
\usepackage{tikzscale} 

\usetikzlibrary{backgrounds,fit,positioning}
\usetikzlibrary{shapes.geometric, arrows}
\tikzstyle{arrow} = [thick,->,>=stealth]
\newcommand{\vect}[1]{\ensuremath{\boldsymbol{\mathrm{#1}}}}

\definecolor{dgreen}{rgb}{0.0, 0.5, 0.0}
\definecolor{dpurple}{rgb}{0.5, 0, 0.6}
\definecolor{dorange}{rgb}{0.9, 0.4, 0}

\usepackage{makecell}   
\usepackage{xcolor}     
\usepackage{array}      
\usepackage{booktabs}   
\usepackage{colortbl}   
\usepackage{multirow}   

\title{ \LARGE \bf 
Intent-Aware MPC for Aircraft Detect-and-Avoid with Response Delay: A Comparative Study with ACAS Xu
} 
\author{Arash Bahari Kordabad, Arabinda Ghosh, Sybert Stroeve, and Sadegh Soudjani
\thanks{Arash Bahari Kordabad, Arabinda Ghosh, and Sadegh Soudjani are with the Max Planck Institute for Software Systems, Kaiserslautern, Germany. E-mail: {\tt\small\{arashbk, arabinda, sadegh\}@mpi-sws.org.} Sybert Stroeve is with the Royal Netherlands Aerospace Centre NLR. E-mail: {\tt\small Sybert.Stroeve@nlr.nl}. \newline This research is supported by the following grants: EIC 101070802 and ERC 101089047.}}

\begin{document}
\bstctlcite{IEEEexample:BSTcontrol}
\maketitle
\thispagestyle{empty}
\pagestyle{empty}

 \begin{abstract} In this paper, we propose an intent-aware Model Predictive Control (MPC) approach for the remain-well-clear (RWC) functionality of a multi-agent aircraft detect-and-avoid (DAA) system and compare its performance with the standardized Airborne Collision Avoidance System Xu (ACAS Xu). The aircraft system is modeled as a linear system for horizontal maneuvering, with advisories on the rate of turn as the control input. Both deterministic and stochastic time delays are considered to account for the lag between control guidance issuance and the response of the aircraft. The capability of the MPC scheme in producing an optimal control profile over the entire horizon is used to mitigate the impact of the delay. We compare the proposed MPC method with ACAS Xu using various evaluation metrics, including loss of DAA well-clear percentage, near mid-air collision percentage, horizontal miss distance, and additional flight distance across different encounter scenarios. It is shown that the MPC scheme achieves better evaluation metrics than ACAS Xu for both deterministic and stochastic scenarios. 
 \end{abstract}
\vspace{0.3cm}
\section{Introduction}
Advanced air mobility (AAM) is a rapidly emerging and new sector of the aerospace industry that aims to safely and efficiently integrate highly automated aircraft in the airspace, including unmanned aircraft systems (UAS) and air taxis. While in conventional (manned) air transport operations, sufficient separation between aircraft is ensured largely by air traffic controllers, in AAM operations more automated approaches for traffic management are needed~\cite{FAA_UTM}. A detect-and-avoid (DAA) system supports a remote pilot of a UAS to observe and avoid nearby air traffic. Its remain-well-clear (RWC) function provides flight path guidance to the remote pilot (RP) so as to maintain an appropriate separation distance. The Airborne Collision Avoidance System Xu (ACAS Xu) is an innovative standardized DAA system to support UAS RPs in remaining well clear and avoiding collisions~\cite{ED275_2020}. The threat resolution module of ACAS Xu uses large lookup tables that were optimized by dynamic programming and a rollout approach for the RWC function~\cite{owen2019acas}. In support of systematic evaluation of ACAS Xu, an agent-based modeling and simulation was developed, which systematically describes interactions between DAA equipped aircraft, RP response delays, and sensor errors~\cite{stroeve2020modeling, Stroeve_Villanueva-Cañizares_Dean_2024}. \emph{Livelock conditions} were found to exist in ACAS Xu supported operations, where the aircraft attain continuing fluctuations away from an intruder and back to their course \textit{without} reaching their destination. It was recognized in~\cite{Stroeve_Villanueva-Cañizares_Dean_2024} that current DAA systems only provide guidance and advisories for avoiding other traffic, which are based on state data. However, they do not take into account planned routes or destinations (intent data) in the guidance provision. As a way forward, this study investigates the added value of an optimization-based DAA system that includes such \emph{intent data}.                       

Model Predictive Control (MPC) has emerged as a promising candidate for addressing these challenges. Rooted in the principle of receding horizon control, MPC optimizes future control actions over a finite prediction horizon while explicitly handling system dynamics and operational constraints. The seminal work by Mayne et al.~\cite{mayne2000constrained} and subsequent developments by Camacho and Bordons~\cite{camacho2007constrained} have established MPC as a powerful tool in both industrial and aerospace applications. Unlike traditional methods, MPC’s ability to incorporate forecasted system behavior and constraints in real time makes it particularly attractive for safety-critical applications such as aircraft detect-and-avoid.

Recent research efforts have further extended the potential of MPC by integrating intent-aware features. The paper~\cite{xu2024intent} introduces a framework that integrates real-time intent prediction with MPC for unmanned aircraft navigation in dynamic environments. An intent prediction module forecasts the trajectories of nearby agents and incorporates these predictions into the MPC formulation, enabling proactive flight path optimization and improved collision avoidance. The paper~\cite{qi2024avoiding} investigates deadlocks in air traffic operations, showing that deadlock avoidance alone fails to address the cascading effects that lead to operational bottlenecks. It develops a resolution framework that integrates advanced scheduling and real-time decision-making strategies to actively resolve blockages.

 Intent-aware controllers seek to predict the future trajectories of nearby aircraft based on their current state and expected maneuvers, modeled as a series of waypoints that the nearby aircraft will pass through. This predictive capability allows the control system to proactively plan avoidance strategies, rather than merely reacting to imminent conflicts. A recent study has demonstrated that when intent information is fused with the predictive power of MPC, yielding smoother and more efficient avoidance maneuvers compared to purely reactive methods~\cite{kordabad2024robust}. Furthermore,~\cite{kordabad2024robust} has proposed MPC as an efficient approach for incorporating such information online. However, despite these promising developments, there remains a gap in the literature regarding a direct performance comparison between intent-aware MPC strategies and established systems such as ACAS Xu using well-known evaluation metrics, particularly under more realistic conditions, such as response delays.

\begin{figure}[ht!]
\centering
\includegraphics[width=0.48\textwidth]{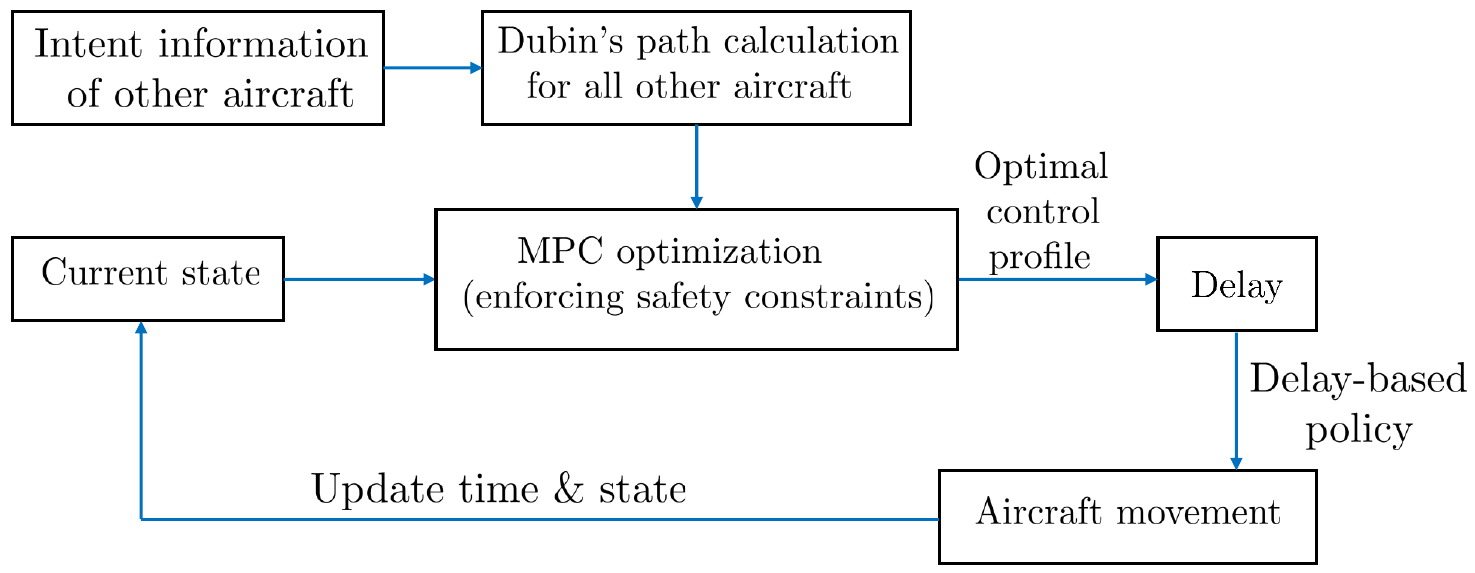}
\caption{
System diagram of the proposed intent-aware MPC framework for multi-aircraft system with response delay for a given aircraft. Intent information is provided as the target state of surrounding aircraft. A Dubins path is used to generate the connecting trajectory. Then the MPC scheme computes the optimal control input sequence, which is used to construct a delay-based policy. This procedure is repeated at each time step.}
\label{fig:100}
\end{figure}

In this paper, we propose an intent-aware MPC framework specifically tailored for aircraft detect-and-avoid. The proposed approach incorporates the intent of surrounding aircraft to improve decision-making and enable proactive maneuver planning. The optimal Dubins path is used to generate a complete path between the current and intended states. Response delay is included in the system modeling, and a delay-based MPC policy is designed to exploit its predictive capabilities for smoother trajectory generation. By systematically comparing the performance of our MPC-based method with the ACAS Xu standard, we aim to evaluate the benefits of integrating intent awareness into MPC schemes. An overview of the proposed intent-aware MPC for delayed multi-aircraft system is illustrated in Figure~\ref{fig:100}. 

The remainder of this paper is organized as follows. Section~\ref{sec:background} reviews the background of the aircraft model, response delay and sensor error. Section~\ref{sec:intentMPC} details the design and implementation of the proposed intent-aware MPC framework for the delayed system. Section~\ref{sec:sim} presents simulation results and comparative analysis against ACAS Xu. Finally, Section~\ref{sec:conc} concludes the paper and outlines directions for future research.

\section{Background on Modeling}
\label{sec:background}
\subsection{Equations of Motion}
\begin{figure}[ht!]
\centering
\includegraphics[width=0.3\textwidth]{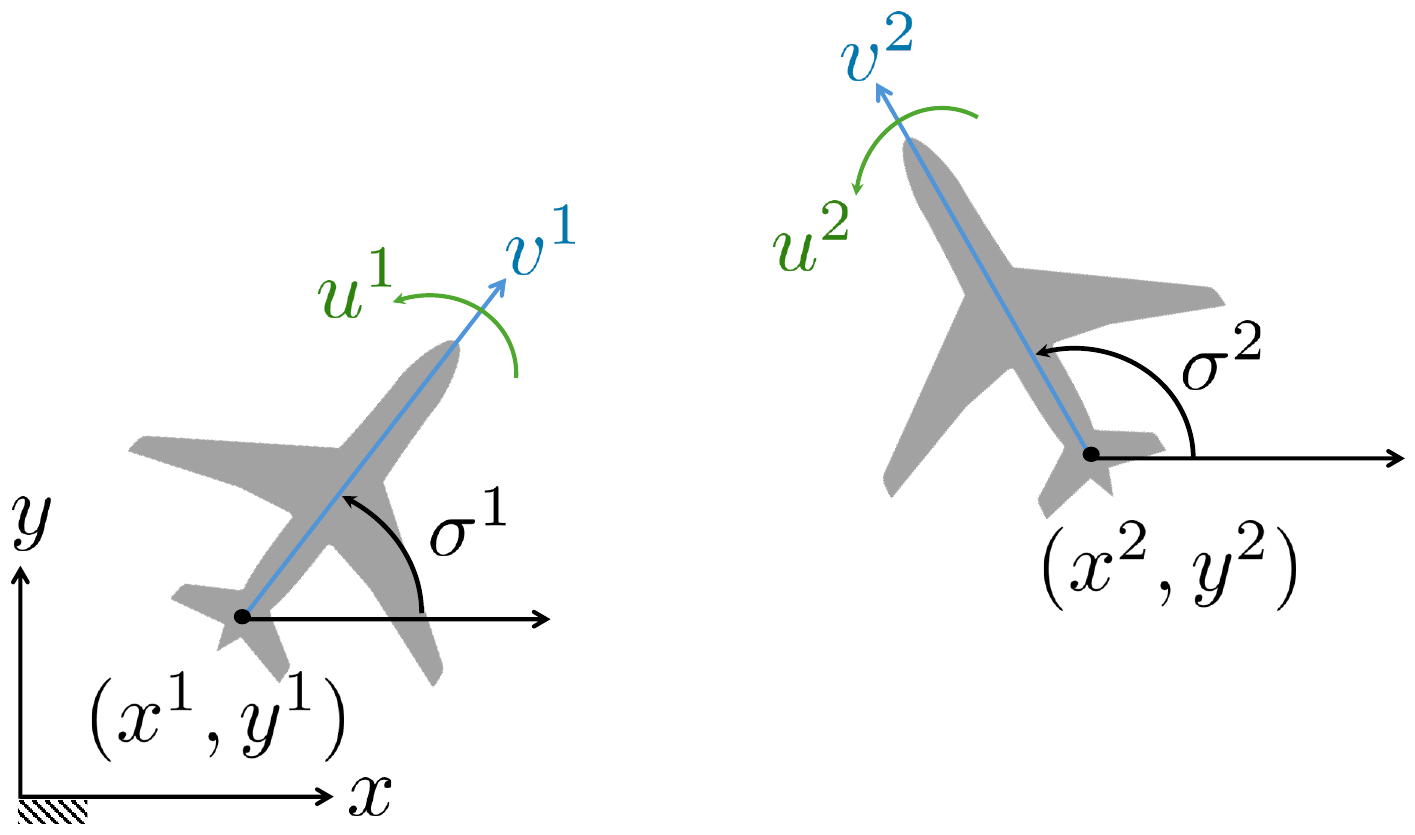}\caption{Geometry of two aircraft in the horizontal plane in the earth-fixed coordinate system. The black variables are the state variables and the greens are the angular velocities as the control inputs. The linear velocities, shown in blue, are constant.}
\label{fig:geo}
\vspace{-10pt}
\end{figure}

We consider a discrete-time model for $n$ aircraft. The equation of motion for each aircraft $i$ is given by:
\begin{equation*}
\vect s_{t+1}^i= \vect s^{i}_t+t_e
    \begin{bmatrix}
        v^{i} \cos{\sigma^{i}_{t}} \\ v^{i} \sin{\sigma^{i}_{t}} \\ u^{i}_t 
    \end{bmatrix},
\end{equation*}
where, the superscript $i\in\{1,2,\ldots, n\}$ represents the $i^{\text{th}}$ aircraft, $\vect s_t^i:=[{x}^{i}_{t}, {y}^{i}_{t}, {\sigma}^{i}_{t}]^\top$ is the state vector of the aircraft $i$ at time $t$, ${x}^{i}_{t}$ and ${y}^{i}_{t}$ are the position states, and ${\sigma}^{i}_{t}$ is the heading angle. The constant $t_e$ is the sampling time and is assumed to be $1$ sec, and $v^{i}$ is the linear velocity that is assumed to be a constant value and $u^{i}_t$ is the angular velocity, which is considered as a control input. In aircraft control, a single control action for angular velocity is often used to reduce the action space and computational complexity, particularly in dynamic programming~\cite{julian2019guaranteeing}. In contrast, the MPC approach can incorporate both linear and angular velocities, allowing for a broader feasible optimization domain~\cite{kordabad2024robust}. However, to ensure a fair comparison with the ACAS Xu approach, this paper considers only angular velocity as the control input. Figure~\ref{fig:geo} shows the geometrical representation of two aircraft in horizontal plane in the earth-fixed coordinate system. 
\subsection{Modeling of Response Delay}\label{sec:back:delay}
In remotely piloted aircraft there is an inherent lag between the provision of an advisory and resulting aircraft maneuver. This delay arises due to several factors, including communication latency and RP reaction time, and it can significantly affect the  performance of the system if not properly managed. Such delays must be accounted for in the control design to ensure reliability. In this study, both deterministic and stochastic delay models are considered. 

\subsubsection{Deterministic Delay Model}
In the deterministic case, a fixed delay is assumed for different scenarios:
\begin{itemize}
    \item \textbf{Automatic:} This represents the situation of an automatic response by the flight management system of the unmanned aircraft. For this a constant delay of $\delta=1$s is set for each aircraft in the encounter-scenario.
    \item \textbf{Quick:} This represents the situation of a quick response by a RP, including the time for downlink, swift reaction of the RP, and uplink. For this a constant delay of $\delta=4$s is set for each aircraft in the encounter-scenario.
    \item \textbf{Slow:} This represents a slower response by the RP, including the time for downlink, reaction of the RP who possibly interacts with an air traffic controller, and uplink. For this a constant delay of $\delta=12$s is set for each aircraft in the encounter-scenario.
\end{itemize}
\subsubsection{Stochastic Delay Model}
The delay of an aircraft is chosen from a lognormal probability distribution $\delta\sim f(x; \mu, \sigma)$ with mean  $\mu=4$s and standard deviation $\sigma=2.5$s:
\begin{equation*}
    f(x; \mu, \sigma) = \begin{cases} 
        0, & x \leq  0, \\
        \frac{1}{\sigma_n x \sqrt{2\pi}} \exp\left(-\frac{(\ln x - \mu_n)^2}{2\sigma_n^2}\right), & x > 0,
    \end{cases}
\end{equation*}
where
\begin{equation*}
    \mu_n = \ln\left(\frac{\mu^2}{\sqrt{\mu^2 + 1}}\right), \quad \sigma_n = \sqrt{\ln\left(\frac{\sigma^2}{\sqrt{\mu^2 + 1}}\right)}.
\end{equation*}
With these parameter values the probability of a delay less than $1$s is $0.0167$ and the probability of a delay more than $12$s is $0.0139$. Thus, the stochastic delays are largely between the settings considered in the deterministic cases.
\subsection{Sensor Errors}
In addition to the uncertainty introduced by stochastic delays, sensor errors in state estimates also affect the control advisories. The position and velocity estimate errors are modeled by first-order autoregressive processes with normally distributed noise (SD is 37.8 m or 4.08 m/s; autocorrelation factors are 0.997)~\cite{stroeve2020modeling}. While the deterministic case assumes perfect sensor readings, the stochastic case accounts for these uncertainties. 

\section{Intent-aware MPC with Response Delay}\label{sec:intentMPC}
Our goal is to design a decentralized control $u^i_t$ for a aircraft $i$ using MPC such that the aircraft approach to its intended \textit{waypoints}, denoted by $\vect s^i_{\text{f}}$ while maintaining a safe separation distance from all other aircraft, i.e., satisfying 
\begin{equation}
    \rho^2\leq (x^i_t-x^j_t)^2+(y^i_t-y^j_t)^2,\, \forall t, \,\,\forall i,j\in\{1,2,\ldots, n\}, \label{eq:const}
\end{equation}
with $i\neq j$, where $\rho$ is the minimum allowed horizontal distance of any two aircraft. 

MPC produces control policies by solving an optimal control problem at each discrete time instant based on the current system state $\vect s$, over a finite, receding horizon. As one of the main benefits, MPC can directly account for state-input constraints in optimization. For the above-mentioned setting, aircraft $i$ solves the following optimization problem
\begin{subequations}
\label{eq:RMPC}
\begin{align}
    \min_{\hat {u}^i_{0:N-1|t},\hat{\vect s}_{0:N|t}^i, \epsilon_{0:N|t}} \,\,\,\, & \|\hat{\vect s}^i_{N|t}-\vect s^i_{\text{f}}\|^2_{Q_{\mathrm{f}}}+\sum_{k=0}^{N-1}  \|\hat{\vect s}^i_{k|t}-\bar{\vect s}^i_k\|^2_Q +\nonumber\\ &\sum_{k=1}^{N-1}\|
    \hat{u}^{i}_{k|t}-\hat{u}^{i}_{k-1|t}\|_R^2+\epsilon^2_k,\label{eq:RMPC:cost}\\
    \mathrm{s.t.} \qquad  &\forall k\in\{0,\ldots, N-1\}:\nonumber \\  &\hat{\vect s}_{k+1|t}^i=  \hat{\vect s}^{i}_{k|t}+t_e
    \begin{bmatrix}
        v^{i} \cos{\hat{\sigma}^{i}_{k|t}} \\ v^{i} \sin{\hat{\sigma}^{i}_{k|t}} \\ \hat{u}^{i}_{k|t} 
    \end{bmatrix}, \label{eq:RMPC:dyn}\\
    &  \rho^2\!-\!\epsilon_{k|t}\leq \! (\hat x^i_{k|t}-\bar x^j_k)^2\!+\!(\hat y^i_{k|t}-\bar y^j_k)^2,\nonumber\\ &\qquad\qquad \forall j\in\{1,\ldots, M\}\setminus \{i\},\label{eq:RMPC:safe} \\ & 0 \leq \epsilon_{k|t}, \\ 
  & \underline{u}^{i}\leq \hat{u}^{i}_{k|t}\leq \bar{u}^{i},\label{eq:RMPC:own}\\
 &\hat{\vect s}^i_{0|t}=\vect s^i_t,\label{eq:RMPC:ini}
\end{align}
\end{subequations}
recursively at each state $\vect s^i_t$ and produces a complete profile of control inputs $\hat{u}^{i,\star}_{\cdot|t} = \{\hat{u}_{0|t}^{i,\star},\ldots, \hat{u}_{N-1|t}^{i,\star}\}$ and corresponding state predictions $\hat{\vect s}^{i,\star}_{\cdot|t}= \{\hat{\vect s}_{0|t}^{i,\star},\ldots, \hat{\vect s}_{N|t}^{i,\star}\}$. The notation \textit{star} $\cdot ^\star$ is used to refer to the optimal value of the decision variables. In order to distinguish between the actual system trajectory and the predicted state-input profile, we use the notation \textit{hat} $\hat \cdot$ for the latter. MPC~\eqref{eq:RMPC} for aircraft $i$ consists of decision variables $\hat {u}^i_{0:N-1|t},\,\hat{\vect s}_{0:N|t}^i$ and slack variables $\vect \epsilon$ with parameters $\{\vect s^i_{\text{f}}, \bar{\vect s}^{1: n}_{0:N-1}, \vect s^i_t\}$. Moreover, $\underline{u}^{i}$ and $\bar{u}^{i}$ represents the lower bound and upper bound for the control input. The cost function in \eqref{eq:RMPC} minimizes the difference of the predicted trajectory $\hat{\vect s}_{0:N|t}^i$ with a reference trajectory $\bar{\vect s}_{0:N-1}^i$ in the sense of the weighted norm defined as $\|\vect x\|^2_{Q}=\vect x^\top Q \vect x$ for any vector $\vect x$ and positive definite matrix $Q$ along with the control deviations. Constraints \eqref{eq:RMPC:dyn}, \eqref{eq:RMPC:safe} and \eqref{eq:RMPC:own} respect the dynamics, RWC  and control input limits. The reference trajectory of aircraft $i$, as well as those of other aircraft required in \eqref{eq:RMPC:safe}, is obtained using intent-based Dubins paths~\cite{kordabad2024robust}, i.e., the complete reference trajectories of other aircraft are assumed to be generated using Dubins paths and are provided as input to the optimization.

Note that due to various uncertainties such as sensor error and mismatch of the actual paths taken by other aircraft with the assumed path based on the Dubin path, the MPC may not be able to find the feasible solution at some instances for hard safety constraint. To resolve this problem, we propose replacing the hard constraint~\eqref{eq:const} with a soft constraint \eqref{eq:RMPC:safe}. It is worth noting that, when only one aircraft is equipped with the DAA system and the other aircraft follow their predefined paths without sensor errors or stochastic delays, the slack variables take a value of zero. This is because the other aircraft exactly follow their predicted Dubins paths, and no additional sources of uncertainty are present. In other scenarios, the slack variables may also be zero, depending on whether the system can enforce the desired safety threshold.

\subsection{Intent-based MPC}
In order to solve MPC~\eqref{eq:RMPC} for a given aircraft $i$ one has to feed $\vect s_{\mathrm{f}}^i$ and $\bar{\vect s}_k^j, \forall j\in\{1,\ldots, n\}$ and $\forall k\in \{0, \ldots, N-1\}$ as parameters to the optimization. Target state $\vect s_{\mathrm{f}}^i$ is given for any aircraft $i$. However, to estimate $\bar{\vect s}_k^j$ for all aircraft $\forall j\in\{1,\ldots, n\}$, we use intent information of other aircraft. Intent is modeled as waypoints of other aircraft given by $\vect s_{\mathrm{f}}^j$  $\forall j\in\{1,\ldots, n\}$ that is available for any aircraft $i$. To use these waypoints $\vect s_{\mathrm{f}}^j$ to generate the entire predicted path $\bar{\vect s}_k^j, \forall k\in \{0, \ldots, N-1\}$ for all other aircraft, we use Dubins path.
 
\begin{figure}[ht!]
   \hspace{0cm}
   { 
        \scalebox{0.8}{ 
            \usetikzlibrary{backgrounds, fit, positioning}
\usetikzlibrary{shapes.geometric, arrows}
\tikzstyle{arrow} = [thick,->,>=stealth]
        \begin{tikzpicture}
            \fill [red!10] (4+1.414,4-1.414) -- (4,4) arc(135:149.63:2) -- cycle;
            \fill [red!10] (0,2) -- (0,0) arc(-90:-37.14:2) -- cycle;    
            \draw [red, dotted, line width=0.2mm] (0,0) arc (-90:270:2);
            \draw [red, dotted, line width=0.2mm] (4,4) arc (135:495:2);
            \draw [red, line width=0.2mm] (0,0) arc (-90:-36.55:2);
            \draw [red, line width=0.2mm] (4,4) arc (135:149.63:2);
            \node at (0, 2) [circle,fill,inner sep=1pt, black]{};
            \node at (4+1.414,4-1.414) [circle,fill,inner sep=1pt, black]{};
            \node at (4+1.414-1.726,4-1.414+1.011) [circle,fill,inner sep=1.5pt, red]{};
            \draw[line width=0.2mm, red] (1.607,0.809) -- (4+1.414-1.726,4-1.414+1.025);
            \node at (1.607,0.809) [circle,fill,inner sep=1.5pt, red]{};
            \node at (0,0) [circle,fill,inner sep=1pt, black]{};
            \node at (4,4) [circle,fill,inner sep=1pt, black]{};
            \draw[arrow, thick] (0,0) -- (1,0);
            \draw[arrow, thick] (4,4) -- (4.707,4.707);
            \node at (0,-0.25) {{$\vect{s}^j_t$}};
            \node at (3.9,4.3) {{$\vect{s}^j_{\mathrm{f}}$}};
        \end{tikzpicture}
        }
    }
    \caption{An LSR Dubins path for aircraft $j$}
    \label{fig:dubin}
\end{figure}
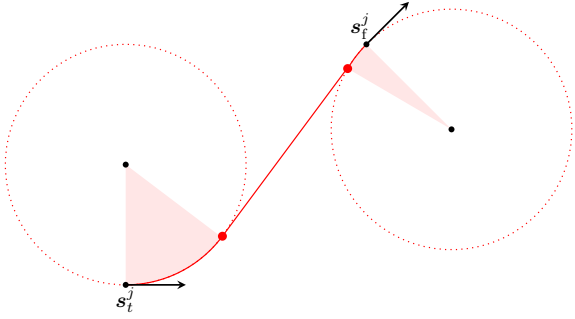

The Dubins path is tangent to the initial conditions (position and direction) and the next waypoint. It determines the path by specifying maneuvers such as turning left (L) by a certain angle, moving straight (S) for a specific duration, and finally turning right (R) by a certain angle. Figure~\ref{fig:dubin} illustrates an example of such a path, denoted as LSR. The red dashed-line circles represent the maximum possible curvature based on the limitations of linear and angular speed. Both geometric~\cite{anisi2003optimal} and analytical methods~\cite{bui1994shortest} are available to compute this optimal path.
More specifically, the optimal Dubins path provides a mapping from the initial-target state pair to the path connecting these two states, i.e., for aircraft $j$, at time $t$ the predicted path to use in~\eqref{eq:RMPC} is given by the following map: 
\begin{equation*}
    \bar{\vect s}_k^j=D(\vect s^j_t,\vect s^j_{\mathrm{f}},k), \quad 0\leq k<N,
\end{equation*}
where $D$ is a mapping from the current state $\vect s^j_t$, the target state $\vect s^j_{\mathrm{f}}$, and the time index $k$ to the corresponding optimal connecting states $\bar{\vect s}_k^j, 0\leq k<N$, generating the optimal Dubins path. This mapping can be obtained based on different types of optimal paths including RSR, RSL, LSR, LSL, RLR, or LRL. The rate of turn is assumed to be either $-2$, $0$, or $2$ deg/s.

Therefore, the optimization in~\eqref{eq:RMPC} is solved simultaneously for all aircraft based on the observation of the current states of other aircraft and a predicted trajectory derived from Dubins path. Subsequently, the MPC policy is executed for all aircraft, the successor states are obtained, and the optimization process is repeated for the next state. More specifically, to incorporate feedback, only the first element $\hat{u}_0^{i,\star}$ of the input sequence $\hat{u}^{i,\star}$ is applied to the system, a successor state $\vect s^i_{t+1}$ is attained in the next time instance, and the optimization \eqref{eq:RMPC} is solved again for the new state $\vect s^i_{t+1}$.

\subsection{MPC Policy with Response Delay}
As explained in Section~\ref{sec:back:delay}, the aircraft system often has a delay in executing the control command. In traditional approaches, the computed control advisory is simply shifted by the delay and applied at the appropriate time. More specifically, at time $t$, we apply the following policy:
\begin{equation*}
\vect \pi _{\text{common}}(\vect s^i_t)=\hat {u}^{i,\star}_{0|t-\delta},
\end{equation*}
as the common method of applying controller for the delayed systems, where $\delta$ is the delay time.

\begin{figure}[ht!]
\centering
\includegraphics[width=0.48\textwidth]{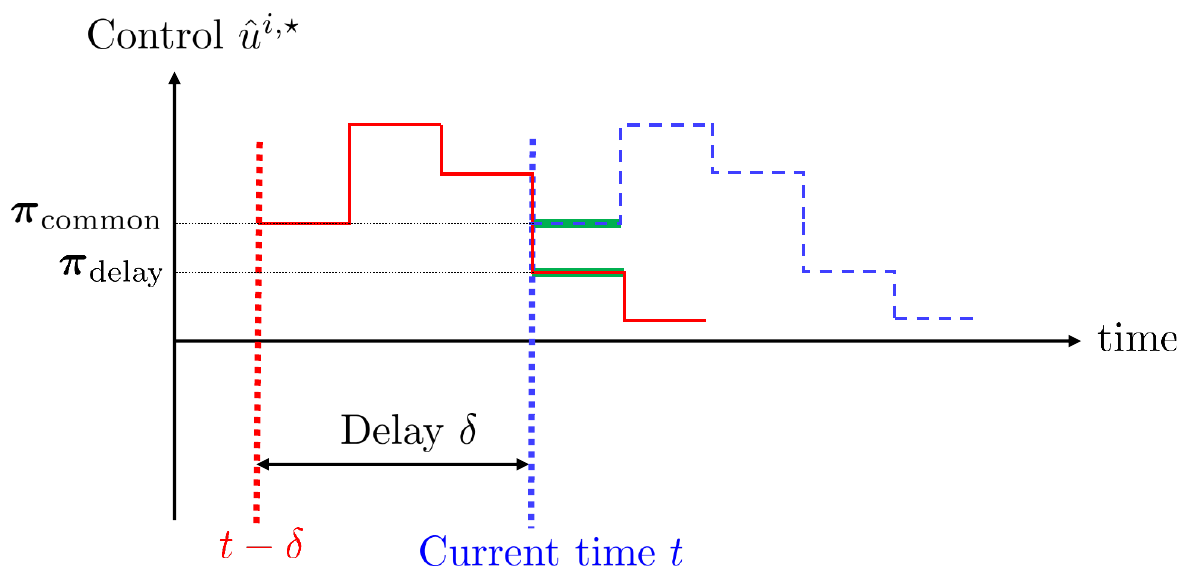}
\caption{Comparison of the proposed MPC scheme policy for delayed systems with common approaches. The blue dotted-line indicates the current time $t$, while the red dotted line represents the shifted time at $t-\delta$, with the corresponding predicted optimal control sequence shown in red. The shifted sequence is shown in blue dashed-line. Green lines represent policies executed using common approaches (simple shifting) or by the proposed MPC policy (aligning with the corresponding predicted element).}
\label{fig:3}
\end{figure}

However, this can result in poor performance and oscillations, as the system may respond suboptimally to outdated control actions.
MPC provides a more advanced approach to effectively address such delays. Instead of delivering a single control action, MPC generates an entire sequence of optimal control actions over a specified prediction horizon. This sequence serves as a comprehensive plan that accounts for future system behavior and constraints. As illustrated in Figure~\ref{fig:3}, rather than merely shifting the control advisory based on the delay (as in common approaches), MPC allows us to align the corresponding control sequence with the system's current state. Specifically, past advisories can be adjusted, and the appropriate element of the sequence can be applied at the current time, ensuring that the control actions are better synchronized with real-time system needs. In this method, at time $t$, we apply the following policy:
\begin{equation*}
\vect \pi _{\text{delay}}(\vect s^i_t)=\hat {u}^{i,\star}_{\delta|t-\delta}.
\end{equation*}
Note that in the stochastic delay case, the delay $\delta$ can be replaced by its mean.  This ability to utilize the control sequence improves system performance by reducing oscillations and enhancing overall stability as demonstrated in the simulation results (section~\ref{sec:sim}). By leveraging the predictive capabilities of MPC, we ensure that the control actions remain relevant and effective, even in the presence of delays. This approach is particularly advantageous in scenarios such as aircraft control, where precision and responsiveness are essential.

\section{Evaluation Metrics and Simulation Results}\label{sec:sim}
In this section, we provide standard performance and safety evaluation metrics and the results of different encounter-scenarios and response delay for two-aircraft system using the common approach for the delay systems and the proposed method utilizing the MPC control profile. 
\subsection{Evaluation of the Performance}
The following metrics are used for evaluation of the proposed MPC scheme in encounter-scenarios.
\begin{itemize}
    \item \textit{Loss of DAA Well Clear (LDWC) percentage}: A loss of DAA Well Clear has been defined to occur for en-route cooperative aircraft~\cite{ED275_2020} if the following three conditions all apply: (1) the projected horizontal miss distance (assuming constant speed) is $\leq4000$ ft ($1219$ m), (2) the so-called modified $\tau$ (for large range approximately the time to pass the range) is less or equal than $35$s, and (3) the vertical separation is less or equal than $450$ft. The LDWC percentage is the part of the runs in a scenario configuration where an LDWC occurred.
    \item \textit{Near Mid-Air Collision (NMAC) percentage}: In the ACAS validation studies traditionally  NMAC events are used as a key metric. It is defined as vertical miss distance (VMD) being $<100$ ft ($30.5$ m) and horizontal miss distance (HMD) being $<500$ ft ($152.4$ m). The NMAC percentage is the part of the runs in a scenario configuration where an NMAC occurred.
    \item \textit{Horizontal Miss Distance (HMD)}: The HMD is the distance in the horizontal plane between a pair of aircraft at the closest point of approach. The mean and standard deviation of the HMD are recorded for each scenario.
    \item  \textit{Additional flight distance (AFD)}: As a result of the DAA advisories the trajectory is adapted and additional distance is traversed. The horizontal distance for each aircraft is determined by the integrals of the traversed distance for the original trajectory and the modified trajectory. The additional distance of an aircraft is the difference of the traversed distances plus the distance between the points at the end of the original and modified trajectories. The additional distance in a run is the sum of the additional distances. The mean and standard deviation of the AFD in the runs of a scenario are gathered.  
\end{itemize}

\subsection{Simulation Results}
We have set $\rho = 1.8\,\text{NM}$, $N = 120$, $Q_{\text{f}}=Q=500I$, $R=1000I$ and all other parameters and models are the same as those used in the ACAS Xu evaluations. The statistics related to the cases with stochastic delays or sensor errors are obtained based on $10$ Monte Carlo simulation runs.

In the considered encounter-scenarios, two aircraft, AC1 and AC2, operate under specific conditions with either only AC2 is equipped with the intent-based MPC strategy or both aircraft are equipped with it. There is no wind, and both aircraft follow planned trajectories that are straight, remain in the horizontal plane, and maintain constant speeds. They fly at the same altitude of $8000$ ft ($2438$ m). All planned aircraft trajectories in an encounter have a HMD of 0 m. AC1 follows a heading of $0$ degrees, while AC2 may have a heading of $45$, $90$, $135$, or $180$ degrees. The speed of AC1 is fixed at $120$ kt ($61.7$ m\slash s), whereas AC2 may travel at either $120$ kt ($61.7$ m\slash s) or $140$ kt ($72$ m\slash s). The maximum allowed turning rate is $2$ deg\slash s for both aircraft. Moreover, different delays have been considered as described in Section~\ref{sec:background}, and in some scenarios, sensor errors have also been incorporated.

\begin{figure}[htbp]
\centering
\includegraphics[width=0.4\textwidth]{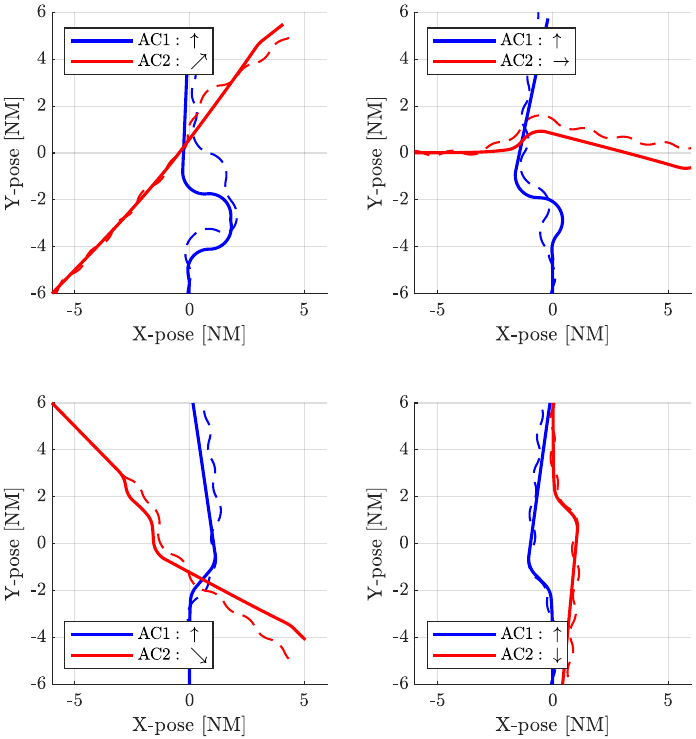}
\caption{Comparison of the MPC policies $\vect\pi_{\mathrm{common}}$ (dashed line) and  $\vect\pi_{\mathrm{delay}}$ (solid line) without sensor errors and for both aircraft equipped, with ``slow" response delay and same speeds for relative headings $45$deg (top-left), $90$deg (top-right), $135$deg (bottom-left), $180$deg (bottom-right).}
\label{fig:0}
\end{figure}

Figure~\ref{fig:0} compares the common MPC policy $\vect\pi_{\mathrm{common}}$ with the proposed intent-aware MPC policy $\vect\pi_{\mathrm{delay}}$. As shown, the proposed method reduces the oscillations caused by system delays by leveraging the prediction capabilities of the MPC scheme tailored for delayed systems. These oscillations are known to arise in such systems, as discussed, for example, in~\cite{Stroeve_Villanueva-Cañizares_Dean_2024}.

\begin{figure}[htbp]
\centering
\includegraphics[width=0.5\textwidth]{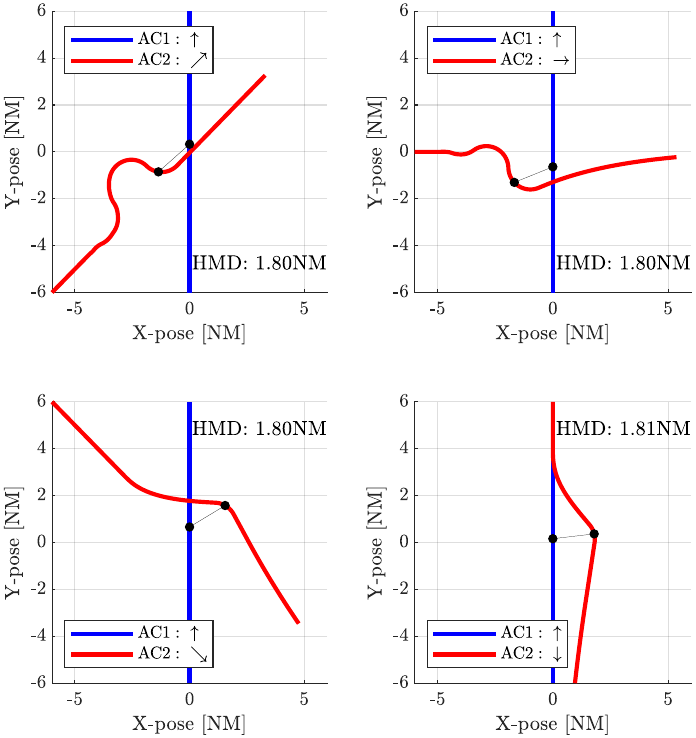}
\caption{Deterministic encounter-scenarios without sensor errors and for only AC2 equipped, ``auto" response delay and same  speeds for relative headings $45$deg (top-left), $90$deg (top-right), $135$deg (bottom-left), $180$deg (bottom-right).}
\label{fig:1}
\end{figure}

\begin{figure}[htbp!]
\centering
\includegraphics[width=0.48\textwidth]{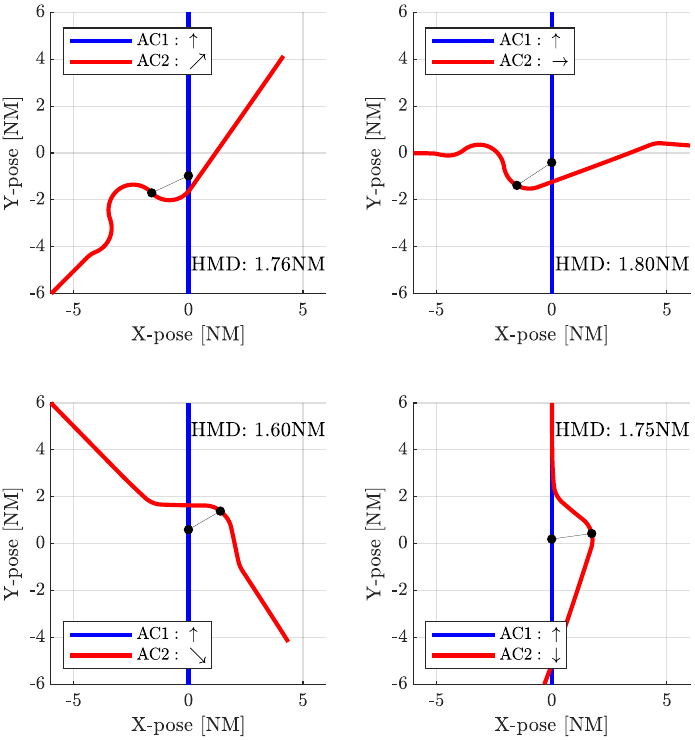}
\caption{Deterministic encounter-scenarios without sensor errors and for only AC2 equipped, ``slow" response delay and different speeds for relative headings $45$deg (top-left), $90$deg (top-right), $135$deg (bottom-left), $180$deg (bottom-right).}
\label{fig:2}
\end{figure}

\begin{figure}[htbp]
\centering
\includegraphics[width=0.48\textwidth]{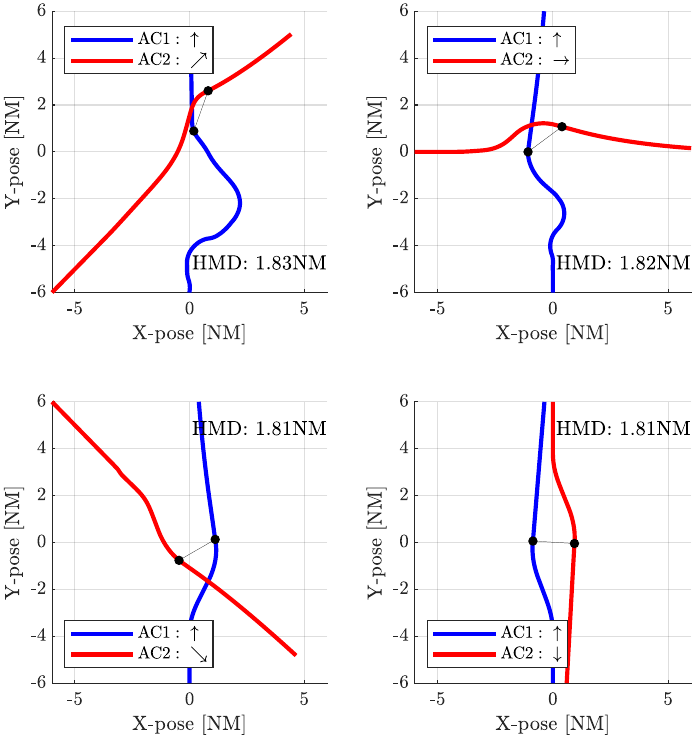}
\caption{Deterministic encounter-scenarios without sensor errors and for both aircraft equipped, ``auto" response delay and same speeds for relative headings $45$deg (top-left), $90$deg (top-right), $135$deg (bottom-left), $180$deg (bottom-right).}
\label{fig:7}
\end{figure}

\begin{figure}[htbp]
\centering
\includegraphics[width=0.48\textwidth]{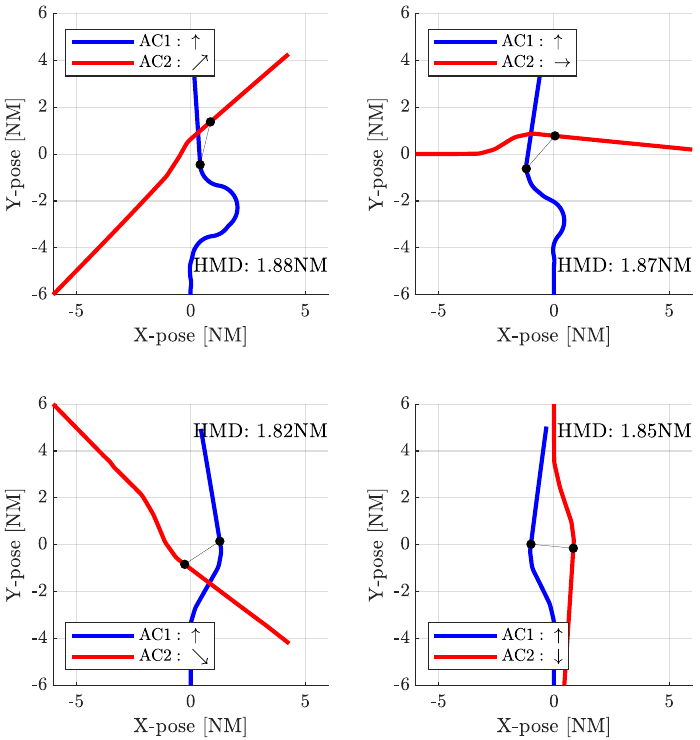}
\caption{Deterministic encounter-scenarios without sensor errors and for both aircraft equipped, ``quick" response delay and different speeds for relative headings $45$deg (top-left), $90$deg (top-right), $135$deg (bottom-left), $180$deg (bottom-right).}
\label{fig:8}
\end{figure}

\begin{figure}[htbp!]
\centering
\includegraphics[width=0.48\textwidth]{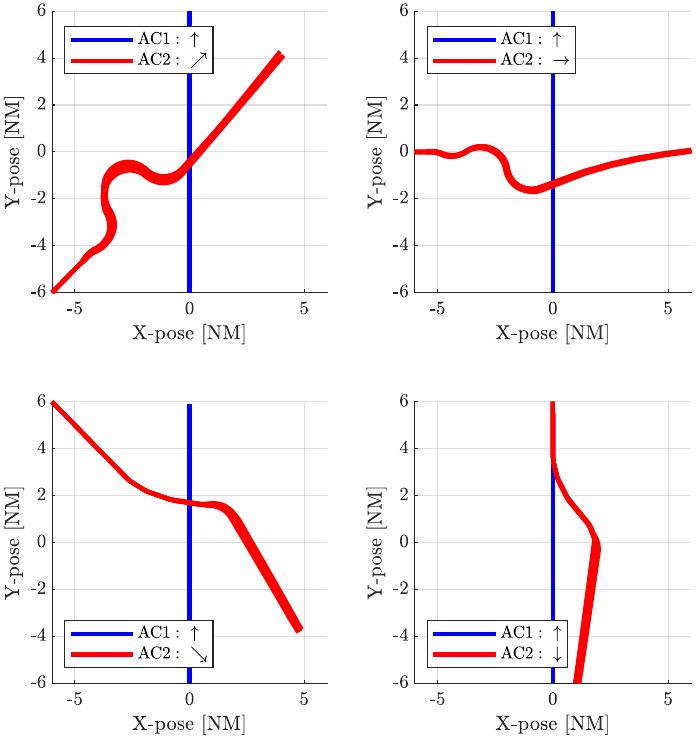}
\caption{Stochastic encounter-scenarios with sensor errors and for only AC2 is equipped, ``quick" response delay and different speeds for relative headings $45$deg (top-left), $90$deg (top-right), $135$deg (bottom-left), $180$deg (bottom-right).}
\label{fig:14}
\end{figure}

\begin{figure}[htbp!]
\centering
\includegraphics[width=0.48\textwidth]{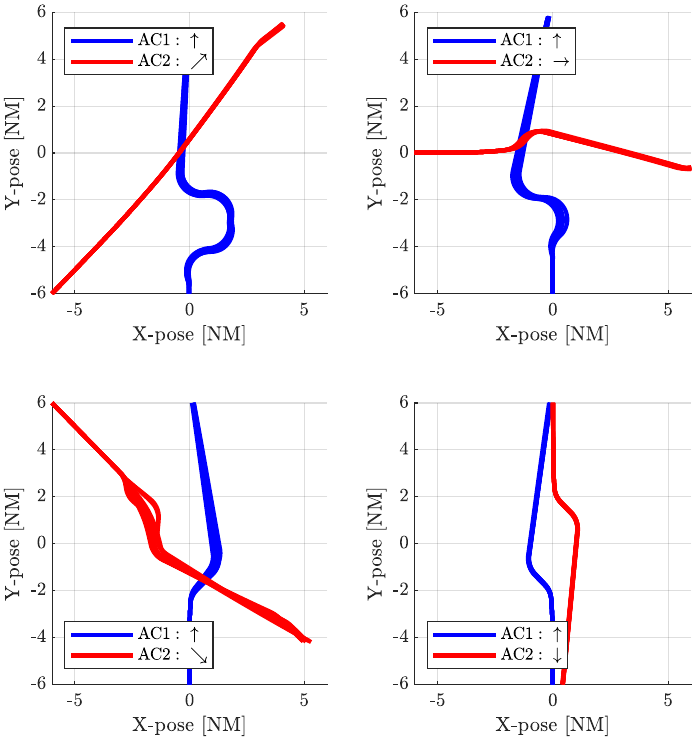}
\caption{Stochastic encounter-scenarios with sensor errors and for both aircraft equipped, ``slow" response delay and same speeds for relative headings $45$deg (top-left), $90$deg (top-right), $135$deg (bottom-left), $180$deg (bottom-right).}
\label{fig:19}
\end{figure}

\begin{figure}[htbp]
\centering
\includegraphics[width=0.48\textwidth]{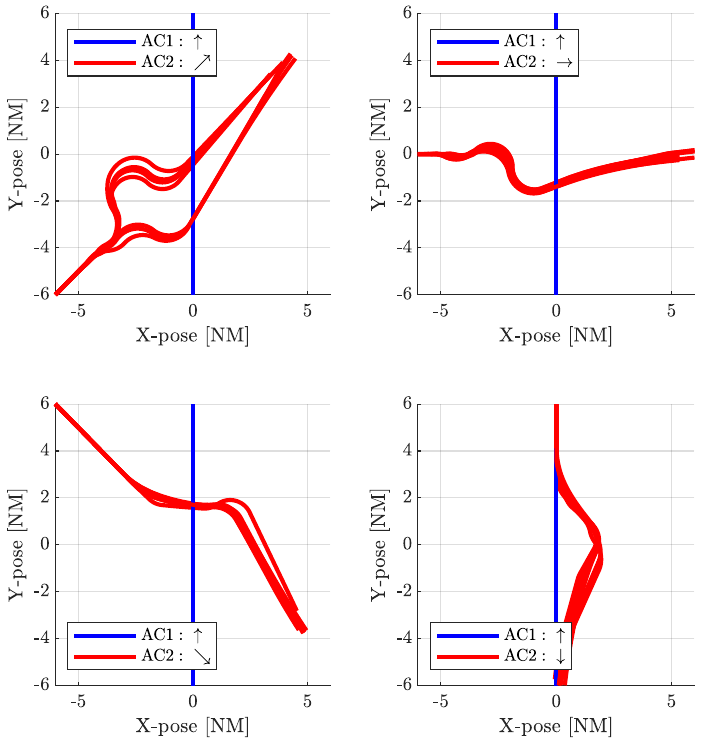}
\caption{Stochastic encounter-scenarios with sensor errors and for only AC2 equipped, ``stochastic" response delay for relative headings $45$deg (top-left), $90$deg (top-right), $135$deg (bottom-left), $180$deg (bottom-right).}
\label{fig:21}
\end{figure}

Figures \ref{fig:1}–\ref{fig:21} present a range of encounter scenarios involving different response delays, relative heading angles, and aircraft speeds, both with and without sensor errors. The scenarios include configurations where only AC2 is equipped with the DAA system as well as those where both aircraft are equipped. Across all scenarios, the proposed intent-aware MPC strategy consistently shows strong performance in resolving conflicts, generating smooth trajectories, and maintaining safe separation—even under significant delays and uncertainty.

In scenarios with longer delays (e.g., Figures~\ref{fig:2} and~\ref{fig:19}), the predictive capability of MPC effectively mitigates the impact of delayed responses and enables timely and efficient avoidance maneuvers. When both aircraft are equipped, mutual intent-awareness facilitates more cooperative behavior, resulting in smaller deviations from planned paths compared to scenarios where only one aircraft is equipped.

Furthermore, sharp turns are largely avoided, which is attributed to the sufficiently long prediction horizon used in the MPC formulation ($N = 120$, corresponding to a $2$-minute lookahead). Overall, the proposed MPC approach demonstrates robustness across all conditions, maintaining safe operation even under high delays and sensor noise regardless of whether only one or both aircraft are equipped with the DAA system.

\begin{figure}[htbp]
\centering
\includegraphics[width=0.48\textwidth]{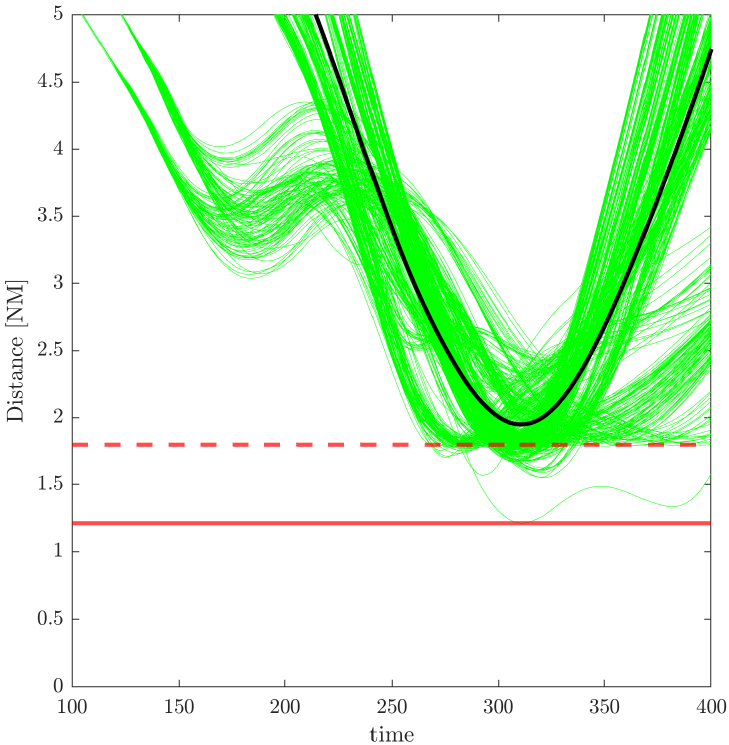}
\caption{The distance of two aircraft for the different encounter-scenarios using the proposed intent-aware MPC approach. The minimum distance among all encounter-scenario is shown in solid red, the soft constraint threshold of MPC is shown in dashed-line and the average is shown in black.}
\label{fig:11}
\end{figure}

Figure~\ref{fig:11} illustrates the distance between two aircraft across various encounter scenarios, along with the minimum distance observed in each case. As shown, despite the presence of response delays and sensor errors, the minimum distance is around $1.2$ NM, demonstrating the effectiveness of the proposed MPC strategy in maintaining safe distances.

\begin{figure}[htbp]
\centering
\includegraphics[width=0.48\textwidth]{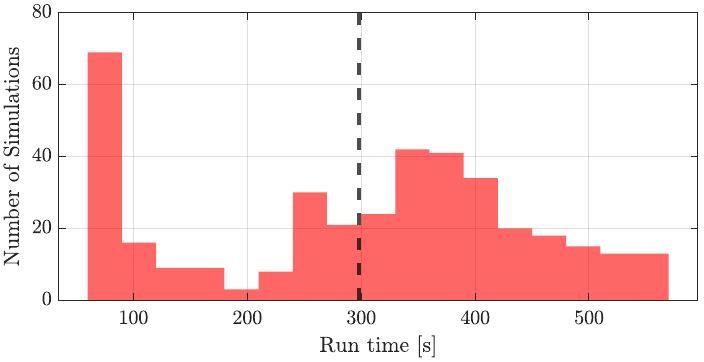}
\caption{Computation time for different encounter-scenarios using the proposed MPC method on a standard laptop.}
\label{fig:rt}
\end{figure}

Figure~\ref{fig:rt} shows the computation time for a single aircraft in the two-aircraft simulation scenario for different encounter-scenarios, providing insight into the computational complexity of the proposed method. The simulations were executed on a standard laptop equipped with an $11$th Gen Intel(R) Core(TM) $i5$-$1145$G7 processor ($2.60$GHz, base frequency $1.50$GHz). As shown, the average computation time is approximately $300$ seconds to simulate $600$ seconds of aircraft operation, indicating that the method is overall suitable for real-time implementation.

The evaluation metrics are summarized in Table~\ref{tab:1}, which show that the MPC approach results in fewer well-clear violations and no NMACs across scenarios, consistently outperforming the ACAS Xu benchmarks. Specifically, across all deterministic and stochastic configurations, the MPC method either eliminates or significantly reduces the rate of LDWC events, and crucially, no NMACs are observed when using MPC. The table also summarizes improvements in HMD and AFD, further validating the effectiveness of the proposed method under various uncertainties.

For benchmarking, ACAS Xu performance was evaluated using the Collision Avoidance Validation and Evaluation Tool (CAVEAT) simulation framework—a stochastic, agent-based tool developed by NLR and everis/NTT-Data for EUROCONTROL~\cite{stroeve2023caveat, stroeve2020modeling,ED275_2020}.

\begin{table}
\caption{\MakeLowercase{Statistics of deterministic and stochastic encounter-scenarios using MPC (blue) and ACAS Xu (black).}}
\label{tab:1}
\centering
\tiny 
\setlength{\tabcolsep}{2pt} 
\renewcommand{\arraystretch}{1.3} 

\setlength{\tabcolsep}{4pt}
\begin{tabular}{|c|c|c|c|c|c|c|c|c|c|}
\hline
\multicolumn{4}{|c|}{\textbf{Scenario}} & \multicolumn{6}{c|}{\textbf{Results}} \\
\cline{1-10}
{Delay} & {Sensor} & {DAA} & {Type} & {LDWC (\%)} & {NMAC (\%)} & \multicolumn{2}{c|}{HMD (m)} & \multicolumn{2}{c|}{AFD (m)} \\
\cline{7-10}
 & {Errors} & {Equipped} & & & & Mean & SD & Mean & SD \\
\hline
 auto  & no  & partial & det. & \makecell{12.50  \% \\\textcolor{blue}{0.000 \%}} & \makecell{0.000  \% \\\textcolor{blue}{0.000 \%}} & \makecell{1905 \\ \textcolor{blue}{3339}} & \makecell{162 \\ \textcolor{blue}{9}} & \makecell{4767 \\ \textcolor{blue}{2133}} & \makecell{4605 \\ \textcolor{blue}{1205}} \\ \hline
 quick & no  & partial & det. & \makecell{12.50  \% \\\textcolor{blue}{0.000 \%}}  & \makecell{0.000  \% \\\textcolor{blue}{0.000 \%}} & \makecell{1923 \\ \textcolor{blue}{3327}} & \makecell{173 \\ \textcolor{blue}{35}} & \makecell{4920  \\ \textcolor{blue}{2217}} & \makecell{4603 \\ \textcolor{blue}{1042}} \\ \hline
 slow  & no  & partial & det. & \makecell{12.50  \% \\\textcolor{blue}{0.000 \%}} & \makecell{0.000  \% \\\textcolor{blue}{0.000 \%}} & \makecell{2048 \\ \textcolor{blue}{3217}} & \makecell{330 \\ \textcolor{blue}{122}} & \makecell{5974 \\ \textcolor{blue}{2520}} & \makecell{4496 \\ \textcolor{blue}{653}} \\ \hline
 auto  & no  & all     & det. & \makecell{12.50  \% \\\textcolor{blue}{0.000 \%}}  & \makecell{0.000  \% \\\textcolor{blue}{0.000 \%}} & \makecell{2357\\ \textcolor{blue}{3368}} & \makecell{267 \\ \textcolor{blue}{22}}  & \makecell{12564 \\ \textcolor{blue}{1763}} & \makecell{11847 \\ \textcolor{blue}{1331}} \\ \hline
 quick & no  & all     & det. & \makecell{0.000  \% \\\textcolor{blue}{0.000 \%}}  & \makecell{0.000  \% \\\textcolor{blue}{0.000 \%}} & \makecell{2183 \\ \textcolor{blue}{3405}} & \makecell{117\\ \textcolor{blue}{47}} & \makecell{12474 \\ \textcolor{blue}{1867}} & \makecell{12026 \\ \textcolor{blue}{1252}} \\ \hline
 slow  & no  & all     & det. & \makecell{25.00  \% \\\textcolor{blue}{0.000 \%}}  & \makecell{12.50  \% \\\textcolor{blue}{0.000 \%}} & \makecell{2292 \\ \textcolor{blue}{3932}} & \makecell{1071 \\ \textcolor{blue}{322}} & \makecell{7544 \\ \textcolor{blue}{2691}} & \makecell{2854 \\ \textcolor{blue}{1231}} \\ \hline
 auto  & yes & partial & stoch. & \makecell{27.00  \% \\\textcolor{blue}{12.50 \%}}  & \makecell{29.25  \% \\\textcolor{blue}{0.000 \%}} & \makecell{2024 \\ \textcolor{blue}{3347}} & \makecell{492\\ \textcolor{blue}{56}} & \makecell{4979 \\ \textcolor{blue}{2221}} & \makecell{4691 \\ \textcolor{blue}{1194}} \\ \hline
 quick & yes & partial & stoch. & \makecell{29.25  \% \\\textcolor{blue}{0.000 \%}} & \makecell{0.000  \% \\\textcolor{blue}{0.000 \%}}& \makecell{2036 \\ \textcolor{blue}{3327}} & \makecell{536 \\ \textcolor{blue}{72}} & \makecell{5300 \\ \textcolor{blue}{2259}} & \makecell{4806 \\ \textcolor{blue}{1044}} \\ \hline
 slow  & yes & partial & stoch. & \makecell{50.63  \% \\\textcolor{blue}{12.50 \%}}  & \makecell{0.250  \% \\\textcolor{blue}{0.000 \%}} & \makecell{1861 \\ \textcolor{blue}{3210}} & \makecell{666 \\ \textcolor{blue}{184}} & \makecell{6244 \\ \textcolor{blue}{2660}} & \makecell{5036 \\ \textcolor{blue}{1082}} \\ \hline
 auto  & yes & all     & stoch. & \makecell{13.50  \% \\\textcolor{blue}{0.000 \%}}  & \makecell{0.125  \% \\\textcolor{blue}{0.000 \%}} & \makecell{2235 \\ \textcolor{blue}{3367}} & \makecell{476 \\ \textcolor{blue}{56}}  & \makecell{12898 \\ \textcolor{blue}{1751}} & \makecell{14352 \\ \textcolor{blue}{1255}} \\ \hline
 quick & yes & all     & stoch. & \makecell{14.75  \% \\\textcolor{blue}{0.000 \%}}  & \makecell{0.375  \% \\\textcolor{blue}{0.000 \%}} & \makecell{2196 \\ \textcolor{blue}{3393}} & \makecell{526 \\ \textcolor{blue}{63}}  & \makecell{12190 \\ \textcolor{blue}{1863}} & \makecell{12873 \\ \textcolor{blue}{1146}} \\ \hline
 slow  & yes & all     & stoch. & \makecell{38.00  \% \\\textcolor{blue}{0.000 \%}}  & \makecell{0.375  \% \\\textcolor{blue}{0.000 \%}} & \makecell{2134 \\ \textcolor{blue}{3962}} & \makecell{729 \\ \textcolor{blue}{286}} & \makecell{10207 \\ \textcolor{blue}{2705}} & \makecell{9376 \\ \textcolor{blue}{1225}} \\ \hline
 stoch & no & partial     & stoch. &   \makecell{15.00  \% \\\textcolor{blue}{0.000\%}} & \makecell{0.000  \% \\\textcolor{blue}{0.000 \%}} & \makecell{1935 \\ \textcolor{blue}{3296}} & \makecell{180 \\ \textcolor{blue}{114}} & \makecell{4971 \\ \textcolor{blue}{2212}} & \makecell{4629 \\ \textcolor{blue}{864}} \\ \hline
 stoch & no & all     & stoch. &   \makecell{10.63  \% \\\textcolor{blue}{0.000 \%}} & \makecell{0.250  \% \\\textcolor{blue}{0.000 \%}} & \makecell{2123 \\ \textcolor{blue}{3396}} & \makecell{257 \\ \textcolor{blue}{46}} & \makecell{11227 \\ \textcolor{blue}{1879}} & \makecell{13894 \\ \textcolor{blue}{1154}}           \\ \hline
 stoch & yes & partial     & stoch. & \makecell{31.25  \% \\\textcolor{blue}{0.000 \%}} & \makecell{0.000  \% \\\textcolor{blue}{0.000 \%}} & \makecell{1996 \\ \textcolor{blue}{3312}} & \makecell{533 \\ \textcolor{blue}{87}} & \makecell{5289 \\ \textcolor{blue}{2150}} & \makecell{4822 \\ \textcolor{blue}{683}}    \\ \hline
 stoch & yes & all     & stoch. & \makecell{16.50  \% \\\textcolor{blue}{0.000 \%}} & \makecell{0.125  \% \\\textcolor{blue}{0.000 \%}} & \makecell{2183 \\ \textcolor{blue}{3531}} & \makecell{526 \\ \textcolor{blue}{209}} & \makecell{11981 \\ \textcolor{blue}{1987}} & \makecell{12826 \\ \textcolor{blue}{1230}} \\
\hline
\end{tabular}
\end{table}

\section{Conclusion}\label{sec:conc}
In this paper, we proposed an intent-aware Model Predictive Control (MPC) approach for multi-aircraft detect-and-avoid and compared its performance with ACAS Xu using standard evaluation metrics. By leveraging intent information and optimizing control actions over a receding horizon, the proposed MPC framework improves decision-making under realistic conditions, including deterministic and stochastic delays as well as sensor errors. Our comparative analysis across various encounter scenarios demonstrated that the proposed MPC method reduces the risk of near mid-air collisions and improves safety metrics (higher well-clear preservation and lower NMAC rates), thereby outperforming ACAS Xu in all tested scenarios. Furthermore, our results highlight the advantage of utilizing the entire predicted control sequence to mitigate delay effects, rather than simple shifting strategies. Future work will focus on extending this approach to more complex air traffic environments (e.g., scenarios with more than two aircraft or varying flight dynamics) and improving computational efficiency for real-time deployment in large-scale multi-agent systems.
\bibliographystyle{IEEEtran}
\bibliography{IntentMPC_ArxivCDC} 
\end{document}